\documentclass[%
 reprint,
 amsmath,amssymb,
 aps,prl
]{revtex4-1}

\usepackage{comment}
\usepackage{graphicx}
\usepackage{tabularx}
\usepackage{amssymb}
\usepackage{amsmath}

\usepackage[utf8]{inputenc}
\usepackage[T2A,T1]{fontenc}
\usepackage[english]{babel}

\usepackage{color}
\usepackage{ulem}
\usepackage{xcolor}
\usepackage{comment}

\usepackage[caption=false]{subfig}

\graphicspath{ {images/} }

\usepackage{amsmath}
\usepackage{amssymb}
\begin{document}
\title{Thickness-independent narrow resonance in a stack of two plasmonic lattices}

\author{Ilia M. Fradkin}
\email{Ilia.Fradkin@skoltech.ru}
\affiliation{Skolkovo Institute of Science and Technology, Nobel Street 3, Moscow 143025, Russia}
\affiliation{Moscow Institute of Physics and Technology, Institutskiy pereulok 9, Moscow Region 141701, Russia}
\author{Sergey A. Dyakov}
\affiliation{Skolkovo Institute of Science and Technology, Nobel Street 3, Moscow 143025, Russia}
\author{Nikolay A. Gippius}
\affiliation{Skolkovo Institute of Science and Technology, Nobel Street 3, Moscow 143025, Russia}
\date{\today}







\begin{abstract}
Plasmonic lattices consisting of nanoparticles in a homogeneous environment are well known for their support of so-called lattice plasmon resonances. They are associated with localized surface plasmons coupled to each other via free propagating photons along the structure.
In this paper, we explore modes in a stack of two identical plasmonic lattices. We demonstrate that such a structure is able to support a mode that is positioned strictly on a Rayleigh anomaly and does not shift with the variation of distance between two lattices in wide limits. Given the fact that period is the most stably reproduced quantity in an experiment, such behavior can be used to simplify the fabrication of structures with resonances at desired energies.
\end{abstract}

\maketitle

\section{Introduction}
Plasmonic lattices form a large class of optical metasurfaces. In recent years they gained great attention due to a combination of a simple for qualitative analysis design and unique optical properties. Hybrid photonic structures with inclusions of such lattices confine light and enhance light-matter interaction due to plasmonic resonances, whereas photonic counterpart provides high-quality resonances. These prerequisites lead to an application of such structures for sensing \cite{Shen2013,rodriguez2011,lodewijks2012,sterl2020design}, light emission enhancement and routing \cite{kolkowski2019lattice,vaskin2019light,rodriguez2012, Guo2015,vecchi2009,ramezani2016modified}, lasing \cite{schokker2017systematic,zhou2013,wang2017,schokker2016lasing}, holography \cite{zheng2015metasurface,ye2016spin,wei2017broadband} and many other purposes.

Plasmonic lattices in homogeneous environments are well-known by specific non-typical lattice plasmon resonances (LPRs). LPRs are the modes associated with collective oscillations of individual nanoparticles coupled to each other by free photons propagating along the lattice. These resonances are unique since they potentially allow us to achieve very narrow lines in a plasmonic structure without the implementation of a photonic waveguide or any other high-$Q$ dielectric resonator. Such resonances have been thoroughly investigated both theoretically and experimentally \cite{guo2017,Augie2008,Rajeeva2018,kravets2008,kravets2018plasmonic,wang2017rich,rodriguez2011,rodriguez2012,Augie2008,Zhao2003,gerasimov2019engineering,zakomirnyi2017refractory,Guo2015,zhou2013}, powerful computational methods that can simplify their consideration have been developed \cite{chen2017general,babicheva2018metasurfaces,babicheva2019analytical}. However, in most studies, only simple lattices have been considered \cite{Zhao2003,rodriguez2012,Garcia2007, gerasimov2019engineering,zakomirnyi2017refractory,rodriguez2011,guo2017,Rajeeva2018,kravets2018plasmonic,wang2017rich,Guo2015,zhou2013}. Recently, lattices with several particles in a cell attracted attention and were considered in a few studies \cite{baur2018, fradkin2020nanoparticle,liu2018light,kolkowski2019lattice,Humphrey2014}. Nevertheless, there are still many promising designs of plasmonic lattices, particularly in homogeneous media, that are of great interest but have not been considered yet.

In this paper, we consider a stack of two plasmonic lattices in a homogeneous environment. We study symmetries of their modes, observe Fabry-Perot resonances, show and explain the coupling of its lattice plasmon resonances and demonstrate that in some cases one of the hybridized LPRs arises exactly on the Rayleigh anomaly~(RA) \cite{maystre2012}. Such mode is narrower than a typical lattice resonance and is observed in a wide range of distances between lattices. Our results show that this effect can be efficiently used to set the resonance energy precisely since it is determined only by the structure period, which is a quantity most stably reproduced in an experiment. Although we do not provide a comprehensive classification of two-lattice structures and their modes, the presented techniques can be further applied to explore any similar structures.

\section{Results and discussion}

\begin{figure}
    \centering
    \includegraphics[width=\linewidth]{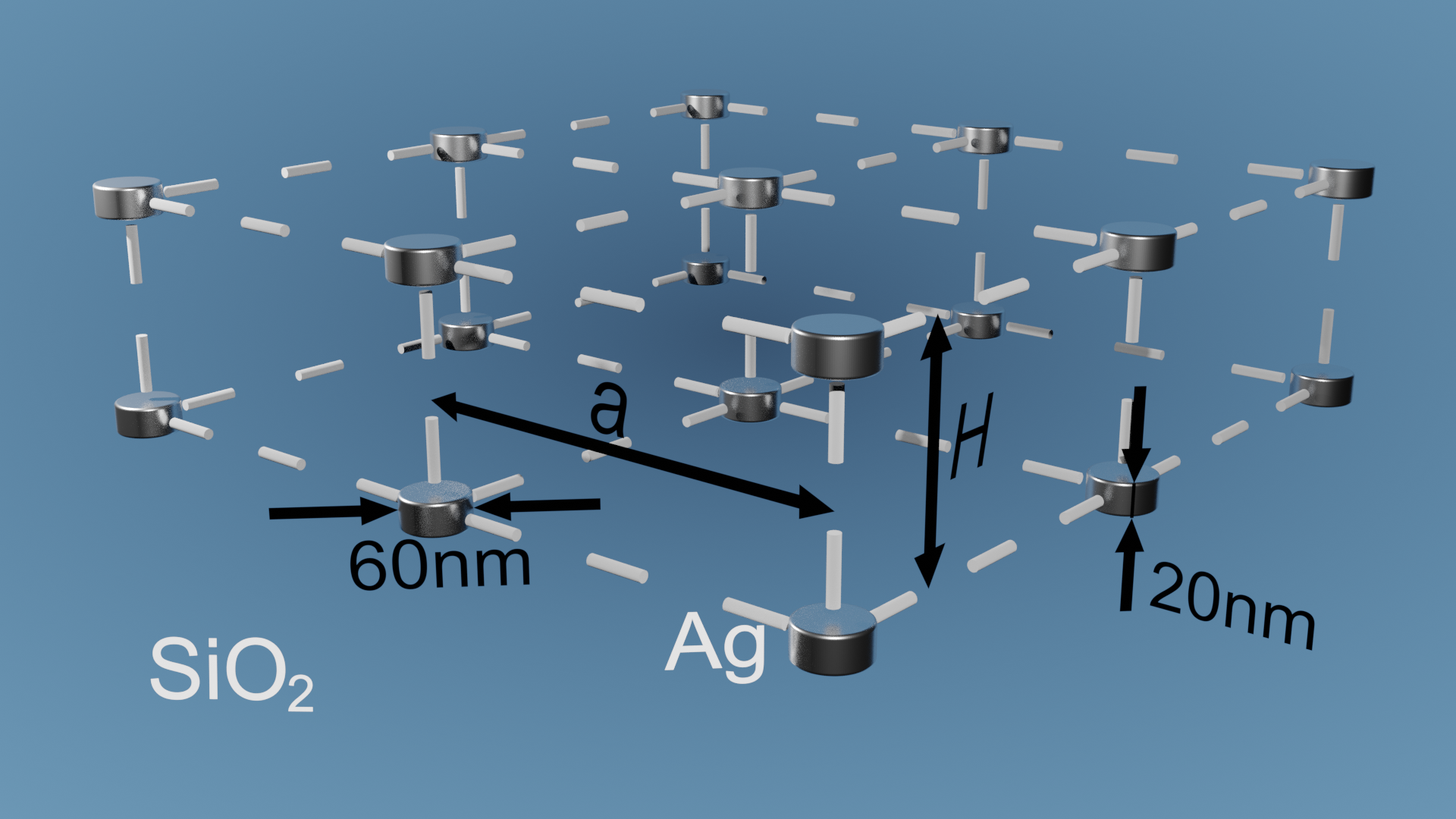}
    \caption{Schematic of the stack of two plasmonic lattices. Each lattice is square and consists of silver nanodiscs. The whole structure is embedded into infinite layer of silica.}
    \label{fig:1}
\end{figure}

Physical phenomena demonstrated in this article do not require a specific choice of materials and particles, so we consider simple square lattices of silver nanodiscs embedded in silica ($\varepsilon_{\mathrm{SiO}_2}=2.1$). Nanodiscs have radii of 30~nm, 20~nm height, and are described by Johnson\&Christy optical constants \cite{JohnsonChristy1972}. Period of the structure, $a$, and thickness, $H$,  (see Fig.~\ref{fig:1}) are varied along the study.

Small plasmonic nanoparticles can be effectively described in the dipole approximation if their dimensions are much smaller than both the distance between them and wavelength of light.
Such an approach has been developed and applied for plasmonic lattices in many papers \cite{schokker2017systematic,chen2017general,reshef2019multiresonant,Augie2008,Rajeeva2018,kravets2018plasmonic,rodriguez2012,reshef2019multiresonant,berkhout2019perfect,Evlyukhin2010,Humphrey2014,Zhao2003,Garcia2007,kolkowski2019lattice,vaskin2019light} and proved itself to be much faster than widespread finite element method (FEM) or finite-difference time-domain (FDTD). Concurrently, its high precision allows us to trust the results and use them for analysis of physical effects as well as for the design of optical devices. In this article all the calculations are conducted via the method developed by our group in recent papers \cite{fradkin2019fourier, fradkin2018fourier,fradkin2020nanoparticle}, which allows to construct a scattering matrix of a plasmonic lattice in dipole approximation \cite{chaumet2003} and integrate it in Fourier modal method (FMM) \cite{tikhodeev2002} also known as Rigorous coupled-wave analysis (RCWA) \cite{moharam1995}. In such representation, the interaction of any number of vertically-spaced lattices is easily accounted for by the calculation of their combined scattering matrix \cite{ko88}.

At the same time, in order to provide an intuitive explanation of the demonstrated effects and interpretation of the results, here we apply the dipole model, which accounts for two spaced lattices as constituents of a single layer. Dipole moments of particles, $\mathbf{P}$, in a certain cell can be found by an application of polarizability tensor, $\hat{\alpha}$, to a background electric field acting on them:
\begin{equation}
\begin{pmatrix}
\mathbf{P}_1 \\ \mathbf{P}_2
\end{pmatrix} = \begin{pmatrix}
\hat{\alpha} &0\\ 0&\hat{\alpha}
\end{pmatrix}\begin{pmatrix}
\mathbf{E}_1^{\mathrm{bg}} \\ \mathbf{E}_2^{\mathrm{bg}}
\end{pmatrix},
\end{equation}
where indices $1,2$ correspond to either upper or lower lattice. Dipole moments of particles in different cells are trivially connected with each other by Bloch's theorem. Taking into account that background field is a sum of an external electric field, $\mathbf{E}^{0}$, of illuminating light and the field rescattered by neighboring particles of both lattices, we obtain the self-consistent system of equations:

\begin{equation}
\begin{pmatrix}
\mathbf{P}_1 \\ \mathbf{P}_2
\end{pmatrix}=
\begin{pmatrix}
\hat{\alpha} &0\\ 0&\hat{\alpha}
\end{pmatrix}\left[
\begin{pmatrix}
\mathbf{E}_1^0 \\ \mathbf{E}_2^0
\end{pmatrix}
+
\begin{pmatrix}\hat{C}_{11}&\hat{C}_{12}\\\hat{C}_{21}&\hat{C}_{22}\end{pmatrix}
\begin{pmatrix}
\mathbf{P}_1 \\ \mathbf{P}_2
\end{pmatrix}\right],
\end{equation}
where $\hat{C}$ blocks are lattice sums of dyadic Green's function \cite{chen2017general,kwadrin2014diffractive} also known as dynamic interaction constants \cite{Belov2005}. Diagonal blocks are associated with the self-action of lattices, and since the background environment is homogeneous they are equal in our case $\hat{C}_{11}=\hat{C}_{22}$. Off-diagonal blocks correspond to an interaction of upper and lower lattices. It should be especially noted that off-diagonal  blocks are not equal to each other, $\hat{C}_{12}\neq\hat{C}_{21}$, and equality is achieved only under some specific conditions, which is discussed in supporting information.

Due to the absence of a relative in-plane shift of two lattices and subsequent high symmetry of the considered structure (see~Fig.~\ref{fig:1})
the off-diagonal elements of all the $\hat{C}$ blocks vanish for normally incident light ($\mathbf{k}_\parallel=0$) (see supporting information). This means that the $x$, $y$, and $z$ -polarized solutions get separated in terms of both dipole moments and incident light. Hence, the $z$ component of the dipole moment can not be excited by normally incident light. Concurrently, $x$ and $y$ -polarized solutions are equivalent to each other which allows us to consider only $x$ polarization without loss of generality:

\begin{multline}
\begin{pmatrix}
P_1^x \\ P_2^x
\end{pmatrix}=
\begin{pmatrix}
\alpha_{xx} &0\\ 0&\alpha_{xx}
\end{pmatrix}\times\\\left[
\begin{pmatrix}
E_1^{0x} \\ E_2^{0x}
\end{pmatrix}
+
\begin{pmatrix}C^{xx}_{11}&C^{xx}_{12}\\C^{xx}_{21}&C^{xx}_{11}\end{pmatrix}
\begin{pmatrix}
P^{x}_1 \\ P^{x}_2
\end{pmatrix}\right].
\end{multline}
The fact that $C_{12}^{xx}=C_{21}^{xx}$ (see supporting information), makes the equations symmetric. In turn, such a symmetric system supports two modes corresponding to in-phase and antiphase dipole moment oscillations, which we denote by A and B subscripts. Further we refer to A and B modes as even and odd correspondingly. Appropriate choice of the basis makes the system diagonal:

\begin{multline}
\begin{pmatrix}
P_\mathrm{A} \\ P_\mathrm{B}
\end{pmatrix}=
\frac{\alpha_{xx}}{\sqrt{2}}\begin{pmatrix}
E_1^{0x}+E_2^{0x} \\ E_1^{0x}-E_2^{0x}
\end{pmatrix}
+\\\alpha_{xx}
\begin{pmatrix}C^{xx}_{11}+C^{xx}_{12}&0\\0&C^{xx}_{11}-C^{xx}_{12}\end{pmatrix}
\begin{pmatrix}
P_\mathrm{A} \\ P_\mathrm{B}
\end{pmatrix},
\end{multline}
where

\begin{align}
\begin{pmatrix}
P_\mathrm{A} \\ P_\mathrm{B}
\end{pmatrix}
&=\frac{1}{\sqrt{2}}\begin{pmatrix}
     1&1\\1&-1
    \end{pmatrix}
    \begin{pmatrix}
P^{x}_1 \\ P^{x}_2
\end{pmatrix}.
\end{align}
Solution of this system gives us amplitudes of the modes in the explicit form:
\begin{align}
    P_\mathrm{A}=\frac{E^{0x}_1(1+e^{ikH})/\sqrt{2}}{\alpha_{xx}^{-1}-(C^{xx}_{11}+C_{12}^{xx})},\label{eqn:dipmom1}\\
    P_\mathrm{B}=\frac{E^{0x}_1(1-e^{ikH})/\sqrt{2}}{\alpha_{xx}^{-1}-(C^{xx}_{11}-C_{12}^{xx})},\label{eqn:dipmom2}
\end{align}
where $k=k_0\sqrt{\varepsilon_{\mathrm{SiO}_2}}$ is the wavenumber in silica and the external incident light is assumed coming from above.
It is important to emphasize that these expressions are strict in the framework of the dipole approximation. However, we derived them not to conduct computations and obtain the precise solution, but for qualitative analysis.

\begin{figure*}
    \centering
    \includegraphics[width=\linewidth]{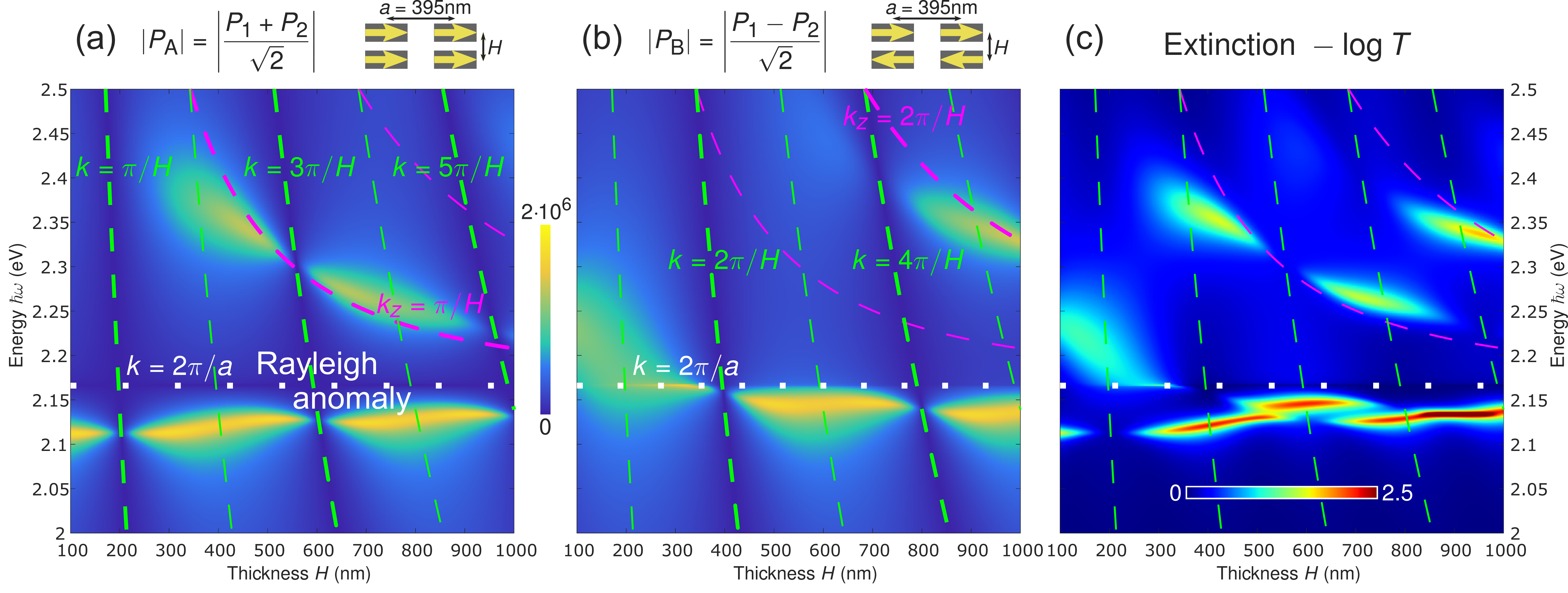}
    \caption{Thickness-dependent spectra of (a) even and (b) odd mode amplitudes. White dotted lines correspond to the energy of RA, thick (or thin) green dashed lines indicate the condition of no-excitation (or maximum excitation)  on panels (a,b). Magenta lines show the estimations of the position of Fabry-Perot modes. In panels (a) and (b) thick lines correspond to modes of the even and odd parities correspondingly. Panel (c) shows the extinction spectrum of the same structure for the normal light incidence.}
    \label{fig:2}
\end{figure*}

As a first example, we consider the structure of $a=395$~nm period and compute its spectra as a function of thickness, $H$. Since ordinary lattice plasmons are observed in close vicinity of Rayleigh anomalies (RA) we consider energies near the first order RAs -  $\left<\pm1,0\right>$ and $\left<0,\pm1\right>$. By means of our approach \cite{fradkin2018fourier,fradkin2019fourier,fradkin2020nanoparticle}, we calculate $\left|P_\mathrm{A}\right|$ and $\left|P_\mathrm{B}\right|$ as they are most convenient for analysis (Fig.~\ref{fig:2}~(a-b)). There are lots of modes and peculiarities in these figures that we explain stepwise below.

In the expressions for the modes amplitudes, (see~Eqns.~\ref{eqn:dipmom1},\ref{eqn:dipmom2}) numerators represent the effectiveness of incident light coupling to corresponding modes. They are periodic functions of $H$ that oscillate between their maxima and zero. From the explicit expressions (Eqns.~\ref{eqn:dipmom1},\ref{eqn:dipmom2}) we derive that amplitude of even mode A is zero for $kH=\pi+2\pi n$ and amplitude of odd mode B for $kH=2\pi n$, $n\in \mathbb{N}$. In other words, when these conditions are satisfied, normally incident light is not able to excite one or another mode because of the symmetry mismatch.
As we can see from Fig.~\ref{fig:2}~(a-b) thick green dashed lines, which correspond to these conditions indeed lay in a dark blue zone of no-excitation. Thin green lines are defined by the same formulas but correspond to the excitation maxima of complementary modes.
When the photon energy matches the level of RA, its wavelength in silica is equal to the period of the structure ($k=2\pi/a$) by definition, which means that the intersection of thick green lines with a white dotted line of RA occurs at $H=a/2+an$ or $H=an$ depending on the symmetry.


As the next step, we consider the hybridization of lattice plasmons. Taking into account the mode's parity and their proximity to RAs, one can easily obtain that the even lattice mode A (or odd lattice mode B) cannot be excited at $H\approx200,600,$ and $1000$~nm (or $H\approx0,400,$ and $800$ nm). However, these modes can be observed at thicknesses in between these values, which is confirmed by Fig.~\ref{fig:2}~(a-b).
In Ref.~\cite{baur2018} it was shown that in the case when both lattices lie in the same plane ($H=0$) and the relative in-plane shift between them tends to zero, $x$-components of diagonal and off-diagonal blocks $C^{xx}$ are equal $C^{xx}_{11}=C_{12}^{xx}$ in close vicinity of the first RA. However, since both of them diverge at RA and this divergence is due to harmonics propagating along the structure ($k_z\approx0$), this relation remains valid for $H>0$ in some limits. Therefore, lattice sums add up in denominator for the A mode amplitude (Eqn.~\ref{eqn:dipmom1}) and we observe the lattice plasmon in Fig.~\ref{fig:2}~(a) when the condition $\mathrm{Re}\alpha_{xx}^{-1} = 2 C^{xx}_{11}$ is satisfied. Also, in this case, the sums diverge on RA and $P_\mathrm{A}$ goes to zero exactly as in the case of a single lattice.

However, much more interesting situation is observed for the odd mode B where the lattice sums cancel each other out, which means that in order to obtain meaningful results we should accurately account for their finite difference and finite contributions of other harmonics.
Here we analyze analytical expressions to explain the most important effects.

\begin{figure*}
    \centering
    \includegraphics[width=\linewidth]{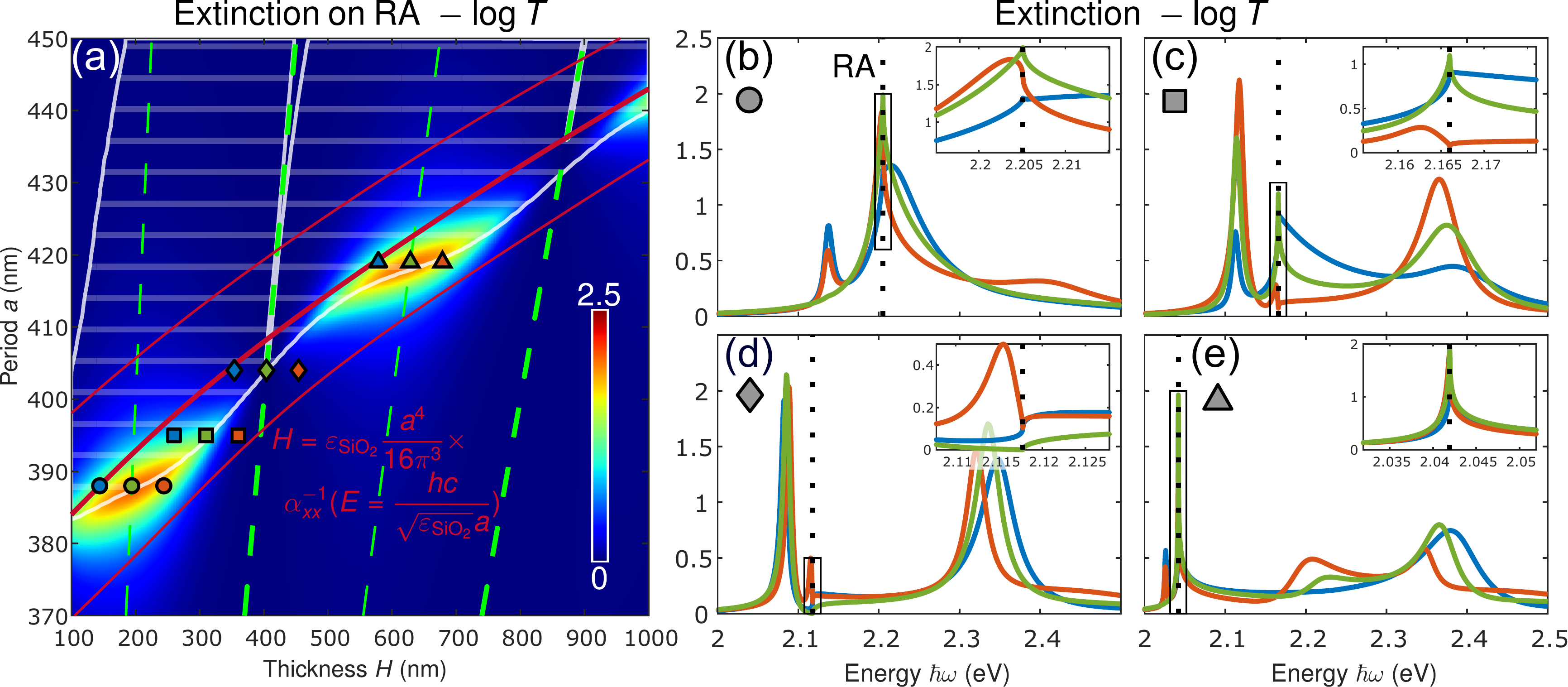}
    \caption{Panel (a) shows the extinction of lattice stack on Rayleigh anomaly as a function of its period and thickness. The crosshatched region with white edge indicates the area in which the peak of the resonance is located exactly on the RA. Red lines show the naive analytical estimation of the resonance ridge position and width, central line corresponds to the real part of the given expression, whereas each of the sidelines is shifted on its imaginary part. Multicolored markers correspond to geometric parameters of lattices considered in panels (b-e). Panels (b-e) show the extinction spectra of the structures defined by markers. The shape of the marker determines the group of the fixed period, whereas colors in each group correspond to different thicknesses. Insets in each panel show shape of lines in close vicinity of RAs.}
    \label{fig:4}
\end{figure*}

It is convenient to represent lattice sums in the vicinity of RA as a sum of diverging term (originating from diffraction harmonics of the first order) and the remaining part: $C^{xx}_{11}=\frac{4\pi}{s}\frac{i k_0^2}{k_z} +\tilde{C}^{xx}_{11}$, $C^{xx}_{12}=\frac{4\pi}{s}\frac{i k_0^2}{k_z}e^{i k_z H} +\tilde{C}^{xx}_{12}(H)$, where $k_z=\sqrt{k^2-(2\pi/a)^2}$. What is important, $\tilde{C}^{xx}_{11}$ is a smooth function of energy near RA and does not depend on $H$ at all, whereas $\tilde{C}^{xx}_{12}$ is smooth as well, but depend on $H$. For relatively large thicknesses (more than 150-200 nm) this dependence is mostly determined by the zero order channel, which forces it to oscillate with a period of the wavelength in silica, which is close to the period of the structure, $a$, near RA.
In this way, denominator can be represented as follows:
\begin{multline}
    \alpha^{-1}_{xx}-C_{11}^{xx}+C_{12}^{xx}=\\ \alpha^{-1}_{xx}+\frac{4\pi}{s}\frac{i k_0^2}{k_z}(e^{i k_z H}-1)-\tilde{C}^{xx}_{11} +\tilde{C}^{xx}_{12}(H)
\end{multline}
The main intrigue comes from the expression $(e^{i k_z H}-1)/k_z$ which is indeterminate on RA, which is easily resolved by an expansion in Taylor series:

\begin{multline}    \alpha^{-1}_{xx}-C_{11}^{xx}+C_{12}^{xx}\approx\\
    \alpha^{-1}_{xx}+\frac{4\pi}{s}\frac{i k_0^2}{k_z}(i k_z H - k_z^2 H^2/2) -\tilde{C}^{xx}_{11} +\tilde{C}^{xx}_{12}(H) = \\    \alpha^{-1}_{xx}- \frac{4\pi k_0^2}{s} H - \frac{4\pi k_0^2}{s} i k_z H^2/2 -\tilde{C}^{xx}_{11} +\tilde{C}^{xx}_{12}(H) \label{eqn:9}
\end{multline}
We see that divergent contributions compensate each other effectively for small thicknesses $H$. The impact of the remaining smooth term $-\tilde{C}^{xx}_{11} +\tilde{C}^{xx}_{12}(H) $ is not enough to change the behaviour of the
structure and, therefore, we observe an ordinary wide line of localized plasmon resonance for $H$ less than 200-250~nm (see Fig.~\ref{fig:2}~(b)). For slightly larger thicknesses, the linear term $\frac{4\pi k_0^2}{s} H$ grows in such a way that at some moment it cancels out $\alpha_{xx}^{-1}$ in the denominator of Eqn.~\ref{eqn:dipmom2} and from $H\approx200$ nm we observe bright yellow line of odd lattice plasmon right on the Rayleigh anomaly. The quadratic term in thickness $- \frac{4\pi k_0^2}{s} i k_z H^2/2$ does not make any contribution when we consider energy of RA ($k_z=0$), however, it indicates that there should be a discontinuity of the energy derivative in this point. Moreover, jump of the derivative grows fast with thickness $H$.
As it will be shown further (see~Fig.~\ref{fig:4}~(b-e)) this leads to strongly non-Lorentzian shapes of resonances. The common effect of the shape violation near RA was already considered for some other structures \cite{akimov2011optical}.
Interestingly, this resonance is much narrower than conventional lattice plasmons. With the further increase of $H$, the resonance condition on RA remains valid until approximately 350~nm, which is very close to the thick green line of no-excitation. When this resonance comes back after the intersection with this line it is shifted to the red zone as it should be in the regime of relatively weak coupling between two lattices. In the limit of a very large thickness, $H$, there is no difference in energy of even, odd, and ordinary LPRs of a single lattice.

The last type of resonance to be discussed is a Fabry-Perot-like one. Indeed, we observe several relatively narrow lines above the RA (see~Fig. \ref{fig:2}~(a,b)). In contrast with LPRs, which are mainly associated with photons having $k_z\approx0$ and no phase difference between two lattices ($\mathrm{Re}k_z H=0$), these modes couple them by some sort of quantized photons. In order to understand their nature, we again analyze the denominators from Eqns.~(\ref{eqn:dipmom1}-\ref{eqn:dipmom2}), but in slightly another way. It is obvious that symmetric and antisymmetric resonances arise when $\alpha^{-1}_{xx}-C_{11}^{xx}\approx\pm C^{xx}_{12}(H)$.
We again assume that interaction 
between lattices is mostly determined by the first diffraction harmonics
$C^{xx}_{12}\propto\frac{i}{k_z}e^{ik_z H}$ and, therefore, $k_z H=-\pi/2 \pm \mathrm{arg} (\alpha^{-1}_{xx}-C_{11}^{xx})+2\pi n$.
An expression $\alpha^{-1}_{xx}-C_{11}^{xx}$ corresponds to a self-action of lattice. If we consider spectrum of the single lattice (see supporting information) then we will see that there is a peak of localized plasmon resonance in the energy range of 2.25-2.35~eV, which means that the phase of the considered expression is approximately $-\pi/2$ in this range. Since $k_x=2\pi/a$, even modes, which are associated with $k_z H\approx \pi +2\pi n$ have the following $H$-dependence $k=\sqrt{ \left({2\pi}/{a}\right)^2+\left({\pi+2 \pi n}/{H}\right)^2}$. For odd modes we have $k_z H\approx 2\pi n$ and $k=\sqrt{\left({2\pi}/{a}\right)^2+\left({2 \pi n}/{H}\right)^2}$ correspondingly. Although these estimations are rough - they are verified by the fact that corresponding thick magenta lines match with rigorously calulated resonances (see~Fig.~\ref{fig:2}~(a,b)).

\begin{figure}
    \centering
    \includegraphics[width=\linewidth]{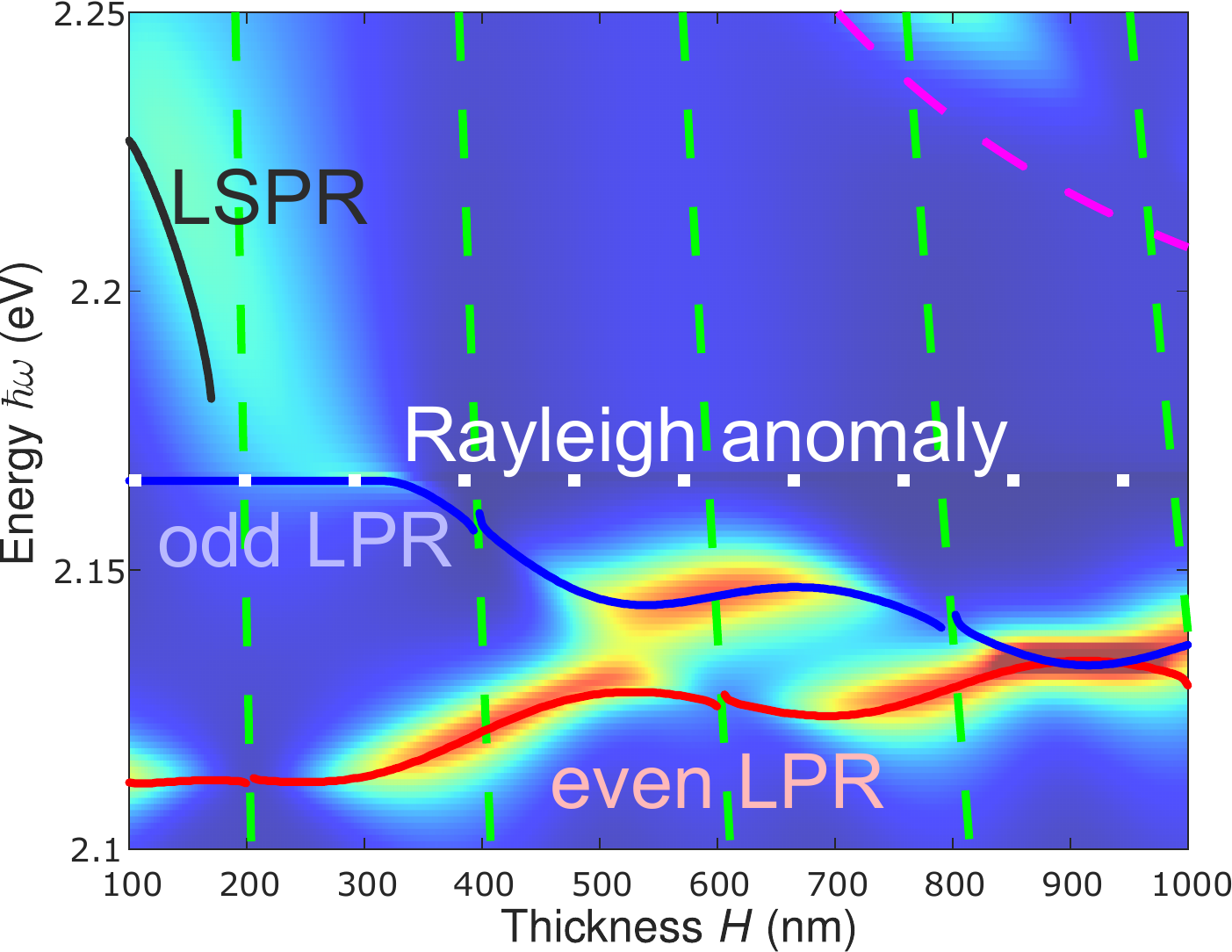}
    \caption{ Thickness dependence of plasmonic modes energy of 395~nm-period lattice. Extinction spectrum is shown in slightly faded colors at the background.}
    \label{fig:3}
\end{figure}

It is worth noting that the first branches of this kind are better to associate with waveguide modes rather than Fabry-Perot ones due to the small phase difference. Also, it is interesting that only the zero phase shift provides us with both even and odd LPR modes.

Amplitudes of even and odd modes are very convenient for analysis, however, in practice, we can observe only integral characteristics that provide us a compound of modes of all parities. For instance, the thickness-dependent extinction spectrum of normally-incident light (Fig.~\ref{fig:2}~(c)) shows that lattice plasmons of complementary symmetries, as well as Fabry-Perot modes, alternate each other with the thickness increase.

We explained the origin of the resonances in this structure, but for further analysis, it is convenient to find energies of resonances as a function of structure parameters. Conventionally it is done by the search of poles of some response function (it would have been scattering matrix in our case) in the plane of complex energy.
However, the most exciting odd (B-type) LPR, which attracts our attention has a non-Lorentzian shape when being located near RA and, therefore, can not be described by a single pole. In this work we do not go into detail of the correct representation of such resonances \cite{akimov2011optical,weiss2017analytical}, but find eigenenergies just as local maxima of $\left|P_\mathrm{A}(E)\right|$ and $\left|P_\mathrm{B}(E)\right|$ functions.

This approach allows us to easily plot the dependence of both LPR energies on the thickness for the same structure. 
From the Fig.~\ref{fig:3} we see how localized surface plasmon resonance (LSPR) gradually transforms into odd lattice plasmon with the thickness increase. Moreover, we see that the peak of $|P_\mathrm{B}|$ stands on RA not only for $H$ in between approximately 200 and 350~nm, but also for smaller thicknesses. Indeed, there is also a small peak, which is not seen by eye, but exists. Blue and red lines of lattice plasmons of complementary parities indeed get closer with the rise of thickness and have the same energy for $H \approx 900$ nm. Also, we clearly see the periodic oscillations of each resonance energy, which are due to the already discussed periodic behavior of $\tilde{C}_{12}^{xx}(H)$.

The presented calculations help us to understand the $H$-dependence of all the modes, including the odd B-type LPR. However, it is also important to explore how this mode behaves with the period variation. In order to do that we scan for periods and thicknesses simultaneously and plot the extinction of corresponding structures on RA (see~Fig.~\ref{fig:4}~(a)). First of all, we see that there is a bright ridge, which corresponds to the resonant conditions of odd LPR. This ridge is cut by the thick green dashed lines $H=a,2a$ that indicate the impossibility to excite odd modes as in Fig.~\ref{fig:2}~(b). The crosshatched region with a white edge in the upper-left part of the graph highlights the area of parameters for which the peak of $\left|P_\mathrm{B}(E)\right|$ is right on the RA. Interestingly, the edge of this region acts as a crest of the ridge and separates the parameter space into two parts in such a way that half of the bright area corresponds to the peak of resonance on RA and the other half to the peak in its close vicinity.

The shape of the bright ridge line can be easily estimated analytically. Indeed, since we consider energy of RA ($E_{\mathrm{RA}}=\frac{h c }{a\sqrt{\varepsilon_{\mathrm{SiO}_2}}}$) then $k_z=0$ and, according to Eqn.~\ref{eqn:9} resonant condition, which connects $H$ and $a$, has the following form:
\begin{equation}
    H=\frac{s}{4\pi k_0^2}\left(\alpha^{-1}_{xx}(E_\mathrm{RA})-\tilde{C}^{xx}_{11}(E_\mathrm{RA},a)+\tilde{C}^{xx}_{12}(E_\mathrm{RA},a,H)\right).
\end{equation}
We have already seen that smooth functions of energy and period $\tilde{C}$ being added to the inverse polarizability tensor does not have significant impact on its behaviour. Therefore, in order to simplify our estimation we just omit them, but note that $\tilde{C}^{xx}_{11}$ has nearly constant complex contribution to the value of $H$ as a function of $a$, whereas $\tilde{C}^{xx}_{12}$ has periodic dependence of $H$ which should make the dependence slightly oscillating.
Taking into account that $s=a^2$ and $k_0^2 = 4\pi^2/a^2/\varepsilon_{\mathrm{SiO}_2}$ in this case, we obtain
\begin{equation}
    H(a)\approx \varepsilon_{\mathrm{SiO}_2}\frac{a^4}{16\pi^3}\alpha^{-1}_{xx}(\frac{hc}{\sqrt{\varepsilon_{\mathrm{SiO}_2}} a}).
\end{equation}
As we see from the Fig.~\ref{fig:4}~(a) red lines, which show both the estimation of the ridge position (real part) as well as it is width (imaginary part) indeed match the precise calculation. Moreover, we see that the discrepancy is well explained by the an unaccounted contribution of secondary terms $\tilde{C}$. 

In order to provide a full picture, we compute extinction spectra for several structures, whose geometric parameters are indicated by markers in Fig~\ref{fig:4}~(a). The first group denoted by circles, which are located at the most bright part of the first "hill" of the ridge, is plotted in Fig~\ref{fig:4}~(b). The main peculiarity of this group is that they illustrate the case in which RA lays right onto the localized resonance and strongly deforms its shape. The green line shows the spectrum of the structure, whose parameters satisfy the relation $H=a/2$, which means that even modes can not be excited and indeed, we observe only single odd resonance. Blue and red lines that have slightly different thicknesses $H$ both demonstrate small peaks of even lattice modes in the red zone. As we can see from the inset all the lines have strongly non-Lorentzian shape and red line in contrast to the other ones have its peak slightly shifted to the red zone, which is in complete agreement with a relative position of the circles with respect to the white line of the ridge.

The next group, depicted by squares, has the period, $a$, of 395 nm, which corresponds to the structures considered in the Figs.~\ref{fig:2}~and~\ref{fig:3}. From Fig.~\ref{fig:4}~(c) we see that in this case even LPRs are much more prominent and besides the two lines of lattice resonances we observe Fabry-Perot modes in the blue zone as well. Also, the extinction of the red line is strongly suppressed at the RA because of the vicinity of the green line of no-excitation.

Diamonds correspond to such a period that odd modes excitation in the vicinity of RA is strongly suppressed. From the inset in Fig~\ref{fig:4}~(d) we see that there is no resonance at all for the green line ($H=a$ in this case). The blue line peak is so small that it is not seen and the peak of the red one is slightly redshifted as it should be. Interestingly, the blue line in the vicinity of the RA has a step-like shape, which might be potentially applied for example in optical filtering.

The last group of triangles corresponds to the second "hill" of the ridge. We see that in this case large thicknesses $H$ result in the observation of several Fabry-Perot modes and very narrow lines of lattice modes due to their strong red shift far from the line of dissipative localized plasmon. 
Similar to the case considered in Fig~\ref{fig:4}~(b), the green line does not demonstrate the peak of the even lattice plasmon due to the specific relation of thickness and period, $H=1.5a$. From the inset, we see that there is almost no significant difference between the lines near the RA. 
What is important, these calculations demonstrate that energy of very narrow resonances can be easily set to $E_{\mathrm{RA}}$ by our approach. We hope that this unique property will be useful in problems, which require setting of high-$Q$ resonances at predetermined positions.

In this paper, we have considered the structure of a rather simple design. However, it has lots of degrees of freedom to be tuned for the achievement of the desired properties. For example, the relative in-plane shift of two lattices can interchange energies of even and odd modes or mix them. The choice of the shape and size of particles as well as the type of the lattice potentially allows to consider resonances with non-trivial polarization of dipole moments, demonstrate chiral and spin-orbit effects. Yet another way to go further in the exploration of plasmonic lattices is an increase of a number of particles in a cell or number of layers. Although it is a trivial step in terms of calculation, it can open exciting opportunities for light manipulation.

\section*{Conclusion}

In this paper, we considered a stack of two plasmonic lattices and analyzed hybridized modes emerging in such structure. We showed that aside from localized plasmon it supports Fabry-Perot modes and lattice plasmons of different symmetries. Particularly, the demonstrated antisymmetric LPR is very untypical. It has a very high $Q$-factor for the given detuning from the line of LSPR.
Moreover, it is located strictly on the Rayleigh anomaly for a wide range of structure parameters, which makes it possible to set the desired energy precisely by an appropriate period and independently choose its thickness from the wide range of accessible values. This unique property may be used either as an additional tuning parameter for optimization of the desired properties or as a protection against experimental errors.
Also, it makes stacks of plasmonic lattices attractive for application in problems for which the precise positioning of resonant lines is crucial. In particular, one can employ them in light emission enhancement and routing, lasing from plasmonic structures, optical filtering.

\section*{Acknowledgements}
This work was supported by the Russian Foundation for Basic Research (Grant No. 18-29-20032).







\bibliographystyle{apsrev4-1}

\end{document}